\def\ra{\rangle}
\def\la{\langle}
\def\be{\begin{equation}}
\def\ee{\end{equation}}
\def\ba{\begin{array}}
\def\ea{\end{array}}

\documentclass[aps,pra,showpacs,showkeys,preprint]{revtex4}
\usepackage{amsmath}
\usepackage{amssymb}
\usepackage{graphicx}
\begin{document}

\title{Separability and Entanglement of Quantum States Based on Covariance Matrices}
\author{Ming Li$^{1}$}
\author{Shao-Ming Fei$^{1,2}$}
\author{Zhi-Xi Wang$^{1}$}
\affiliation{$^1$Department of Mathematics, Capital Normal University, Beijing 100037,
China\\
$^2$Institut f\"ur Angewandte Mathematik, Universit\"at Bonn, D-53115,
Germany}

\begin{abstract}
We investigate the separability of quantum states based on
covariance matrices. Separability criteria are presented for
multipartite states. The lower bound of concurrence proposed in
Phys. Rev. A. 75, 052320 (2007) is improved by optimizing the local
orthonormal observables.
\end{abstract}


\maketitle

As one of the most striking features of quantum phenomena,
quantum entanglement has been identified as a key
non-local resource in quantum information processing such as
quantum computation, quantum teleportation,
dense coding, quantum cryptographic
schemes, entanglement swapping and remote
state preparation \cite{book}. The study of quantum
information processing has spurred a flurry of activities in the
investigation of quantum entanglements. Nevertheless, despite the
potential applications of quantum entangled states, the theory of
quantum entanglement itself is far from being satisfied.
One of the important problems in the theory of quantum
entanglement is the separability: to decide whether or not a given quantum state
is entangled. In principle the problem could be solved by
calculating the measure of entanglement.
However most proposed measures of entanglement
involve extremizations which are difficult to handle analytically.

There have been some (necessary)
criteria for separability, the Bell inequalities \cite{Bell64},
PPT (positive partial transposition) \cite{peres} (which is also
sufficient for the cases $2\times 2$ and $2\times 3$ bipartite
systems \cite{3hPLA223}), reduction criterion
\cite{2hPRA99,cag99}, majorization criterion\cite{nielson01},
entanglement witnesses \cite{3hPLA223,ter00,lkch00},
realignment \cite{ChenQIC03,ru02,chenPLA02} and generalized
realignment \cite{chenkai}, as well as some necessary and
sufficient operational criteria for low rank density matrices
\cite{hlvc00,afg01,feipla02}.

In {\cite{julio}} by using the Bloch representation of density matrices the author has
presented a separability criterion that is independent
of PPT and realignment criteria.
It is also generalized to multipartite case
{\cite{hassan}}. In \cite{hofmann} a criterion based on local
uncertainty relations has been presented. It has been shown
that the criterion based on local uncertainty
relations is strictly stronger than the realignment criterion {\cite{0604050}}.
The covariance matrices are then introduced to solve the separability problem in
\cite{117903,0611282,devi,devi2}. Recently a criterion
which is strictly stronger
than the realignment criterion and its nonlinear entanglement witnesses
introduced in \cite{0604050} has also been presented \cite{zzzg}.

In this letter we study the separability problem by using the
covariance matrix approach. An alternative separability criterion
is obtained for bipartite systems.
The local orthonormal observable dependent lower bound of concurrence
proposed in \cite{0611282} is optimized.
The covariance matrix approach is applied to multipartite systems
and a set of separability criteria is obtained.

We first give a brief review of covariance matrix criterion proposed
in \cite{0611282}. Let ${\mathcal {H}}^{A}_{d}$ and
${\mathcal {H}}^{B}_{d}$ be $d$-dimensional complex vector spaces,
and $\rho_{AB}$ a bipartite quantum state in
${\mathcal {H}}^{A}_{d}\otimes{\mathcal {H}}^{B}_{d}$.
Let $A_{k}$ (resp. $B_{k}$) be $d^{2}$ observables on
${\mathcal{H}}^{A}_{d}$ (resp. ${\mathcal{H}}^{B}_{d}$)
such that they form an orthonormal normalized basis
of the observable space, satisfying
$tr(A_{k}A_{l})=\delta_{k,l}$ (resp. $tr(B_{k}B_{l})=\delta_{k,l}$).
Consider the total set $\{M_{k}\}=\{A_{k}\otimes I, I\otimes B_{k}\}$.
It can be proven that {\cite{0604050}},
\begin{eqnarray}\label{lemma1}
\sum\limits_{k=1}^{N^{2}}(M_{k})^{2}=dI,\quad\quad\quad
\sum\limits_{k=1}^{N^{2}}\langle M_{k} \rangle^{2}=tr(\rho^{2}_{AB}).
\end{eqnarray}

The covariance matrix $\gamma$ is defined with entries
\begin{eqnarray}\label{gammae}
\label{c} \gamma_{ij}(\rho_{AB}, \{M_{k}\})=\frac{\langle
M_{i}M_{j}\rangle+\langle M_{j}M_{i}\rangle}{2}-\langle
M_{i}\rangle\langle M_{j}\rangle,
\end{eqnarray}
which has a block structure \cite{0611282}:
\begin{eqnarray}
\label{def}
\gamma=\left(%
    \begin{array}{cc}
      A & C \\
      C^{T} & B \\
    \end{array}%
    \right),
\end{eqnarray}
where $A=\gamma(\rho_{A},\{A_{k}\}), B=\gamma(\rho_{B},\{B_{k}\}),
C_{ij}=\langle A_{i}\otimes B_{j} \rangle_{\rho_{AB}}-\langle A_{i}
\rangle_{\rho_{A}}\langle B_{j} \rangle_{\rho_{B}}$, $\rho_{A}=Tr_B(\rho_{AB})$,
$\rho_{B}=Tr_A(\rho_{AB})$.
Such covariance matrix has a concavity property: for a mixed density matrix
$\rho=\sum\limits_{k}p_{k}\rho_{k}$ with $p_{k}\geq 0$ and
$\sum\limits_{k}p_{k}=1$, one has $\gamma(\rho)\geq
\sum\limits_{k}p_{k}\gamma(\rho_{k})$.

For a bipartite product state
$\rho_{AB}=\rho_{A}\otimes\rho_{B}$, $C$ in $(\ref{def})$ is zero.
Generally if $\rho_{AB}$ is separable, then there exist
states $|a_{k}\rangle\langle a_{k}|$ on ${\mathcal {H}}^{A}_{d}$,
$|b_{k}\rangle\langle b_{k}|$ on ${\mathcal {H}}^{B}_{d}$ and $p_{k}$ such that
\be\label{p1}
\gamma(\rho)\geq \kappa_{A}\oplus \kappa_{B},
\ee
where
$\kappa_{A}=\sum p_{k}\gamma(|a_{k}\rangle\langle a_{k}|,\{A_{k}\})$,
$\kappa_{B}=\sum p_{k}\gamma(|b_{k}\rangle\langle b_{k}|,\{B_{k}\})$.

The so called covariance matrix criterion (\ref{p1}) is made more
efficient and physically plausible in \cite{0611282}.
For a separable bipartite state, it has been shown that
\begin{eqnarray}\label{p2}
\sum\limits_{i=1}^{d^{2}}|C_{ii}|\leq\frac{
(1-tr(\rho_{A}^{2}))+(1-tr(\rho_{B}^{2}))}{2}.
\end{eqnarray}

Criterion (\ref{p2}) depends on the choice of the orthonormal
normalized basis of observable. In fact the term
$\sum\limits_{i=1}^{d^{2}}|C_{ii}|$ has an upper bound $||C||_{KF}$
which is invariant under
unitary transformation and can be attained by choosing proper local
orthonormal observable basis, where $||C||_{KF}$ stands for the Ky Fan norm of $C$,
$||C||_{KF}=tr\sqrt{CC^{\dag}}$, with $\dag$ denoting the transpose and conjugation.
It has been shown in \cite{zhang} that
if $\rho_{AB}$ is separable, then
\begin{eqnarray}
\label{th2}
||C||_{KF}\leq
\frac{(1-tr(\rho_{A}^{2}))+(1-tr(\rho_{B}^{2}))}{2}.
\end{eqnarray}

From the covariance matrix approach, we can also get an alternative
criterion.
From (\ref{def}) and (\ref{p1}) we have that if $\rho_{AB}$ is
separable, then
\begin{eqnarray}
X\equiv\left(
    \begin{array}{cc}
      A-\kappa_{A} & C \\
      C^{T} & B-\kappa_{B} \\
          \end{array}
    \right)\geq 0.
\end{eqnarray}
Hence all the $2\times 2$ minor submatrices of X must be positive. Namely one has
$$ \left| \begin{array}{cc}
(A-\kappa_{A})_{ii} & C_{ij}\\
C_{ji} & (B-\kappa_{B})_{jj}\\
\end{array}
\right|\geq 0,
$$
i.e.
$(A-\kappa_{A})_{ii}(B-\kappa_{B})_{jj}\geq C_{ij}^{2}$. Summing
over all i, j and using (\ref{lemma1}), we get
\begin{eqnarray*}
\sum\limits_{i,j=1}^{d^{2}}C_{i,j}^{2}&\leq&(tr A-tr
\kappa_{A}))(tr B-tr \kappa_{B})\\
&=&(d-tr(\rho_{A}^{2})-d+1)(d-tr(\rho_{B}^{2})-d+1)
=(1-tr(\rho_{A}^{2}))(1-tr(\rho_{B}^{2})).
\end{eqnarray*}
That is \be\label{th1} ||C||_{HS}^{2}\leq
(1-tr(\rho_{A}^{2}))(1-tr(\rho_{B}^{2})), \ee where $||C||_{HS}$
stands for the Euclid norm of $C$, i.e.
$||C||_{HS}=\sqrt{tr(CC^{\dag})}$.

Formulae (\ref{th2}) and (\ref{th1}) are independent and could be
complement. When
$$
\sqrt{(1-tr(\rho_{A}^{2}))(1-tr(\rho_{B}^{2}))}< ||C||_{HS} \leq
||C||_{KF} \leq
\frac{(1-tr(\rho_{A}^{2}))+(1-tr(\rho_{B}^{2}))}{2},
$$
(\ref{th1}) can recognize the entanglement but (\ref{th2}) can not.
When
$$
||C||_{HS} \leq \sqrt{(1-tr(\rho_{A}^{2}))(1-tr(\rho_{B}^{2}))} \leq
\frac{(1-tr(\rho_{A}^{2}))+(1-tr(\rho_{B}^{2}))}{2} < ||C||_{KF},
$$
(\ref{th2}) can recognize the entanglement while (\ref{th1}) not.

The separability of a quantum state can also be investigated by
computing the concurrence.
The concurrence of a
pure state $|\psi\ra$ is given by $C(|\psi\ra)=\sqrt{2(1-Tr\rho_{A}^{2})}$ \cite{rungta},
where $\rho_{A}=Tr_{B}|\psi\ra\la\psi|$.
Let $\rho$ be a state in
${\mathcal {H}}^{A}_{M}\otimes{\mathcal {H}}^{B}_{N}$, $M\leq N$.
The definition is extended
to general mixed states $\rho$ by the convex roof,
\begin{eqnarray}
C(\rho)=\min\limits_{\{p_{i},|\psi_{i}\ra\}}\left\{\sum_{i}p_{i}C(\psi_{i}):
\rho=\sum_{i}p_{i}|\psi_{i}\ra\la\psi_{i}|\right\}.
\end{eqnarray}
In \cite{chenprl} a lower bound of $C(\rho)$ has been obtained,
\begin{eqnarray}\label{con0}
C(\rho)\geq
\sqrt{\frac{2}{M(M-1)}}\left[Max(||T_{A}(\rho)||,||R(\rho)||)-1\right],
\end{eqnarray}
where $T_A$ and $R$ stand for partial transpose with respect to subsystem $A$
and realignment respectively.
This bound is further improved based on local
uncertainty relations \cite{vicente},
\begin{eqnarray}\label{con}
C(\rho)\geq
\frac{M+N-2-\sum_{i}\triangle_{\rho}^{2}(G_{i}^{A}\otimes I+
I\otimes G_{i}^{B})}{\sqrt{2M(M-1)}},
\end{eqnarray}
where ${G_{i}^{A}}$ and $G_{i}^{B}$ are any set of local orthonormal observables.

Bound (\ref{con}) again depends on the choice of the local
orthonormal observables. In the following we show that this bound
can be also optimized, in the sense that a local orthonormal observable-independent
up bound of the right hand side of (\ref{con}) can be obtained.

{\sf [Theorem 1]} Let $\rho$ be a bipartite state in ${\mathcal {H}}_{M}^{A}\otimes{\mathcal
{H}}_{N}^{B}$. $C(\rho)$ satisfies
\begin{eqnarray}\label{con2}
C(\rho)\geq
\frac{2||C||_{KF}-(1-Tr\rho_{A}^{2})-(1-Tr\rho_{B}^{2})}{\sqrt{2M(M-1)}}.
\end{eqnarray}

{\sf [Proof]} The other orthonormal normalized basis of the local orthonormal
observable space can be obtained from $A_{i}$ and $B_{i}$ by unitary transformations $U$ and
$V$: ${\widetilde{A}}_{i}= \sum\limits_{l}U_{il}A_{l}$ and
${\widetilde{B}}_{j}= \sum\limits_{m}V_{jm}^{*}B_{m}$. Select $U$
and $V$ so that $C=U^{\dag}\Lambda V$ is the singular value
decomposition of $C$. Then the new observables can be
written as ${\widetilde{A}}_{i}= \sum\limits_{l}U_{il}A_{l}$,
${\widetilde{B}}_{j}=-\sum\limits_{m}V_{jm}^{*}B_{m}$. We have
\begin{eqnarray*}
\sum_{i}\triangle_{\rho}^{2}({\widetilde{A}}_{i}\otimes I+
I\otimes {\widetilde{B}}_{i})
&=&\sum_{i}[\triangle_{\rho_{A}}^{2}({\widetilde{A}}_{i})
+\triangle_{\rho_{A}}^{2}({\widetilde{B}}_{i})
+2(\la{\widetilde{A}}_{i}\otimes{\widetilde{B}}_{i}\ra
-\la{\widetilde{A}}_{i}\ra\la{\widetilde{B}}_{i}\ra)]\\
&=&M-Tr\rho_{A}^{2}+N-Tr\rho_{B}^{2}-2\sum_{i}(UCV^{\dag})_{ii}\\
&=&M-Tr\rho_{A}^{2}+N-Tr\rho_{B}^{2}-2||C||_{KF}.
\end{eqnarray*}
Substituting above relation to (\ref{con}) we get (\ref{con2}).
\hfill \rule{1ex}{1ex}

Bound $(\ref{con2})$ does not depend on the choice of local
orthonormal observables. It can be easily applied and realized by
direct measurements in experiments.
It is in accord with the result in \cite{zhang} where optimization of entanglement
witness based on local uncertainty relation has been taken into account.
As an example let us consider
the $3\times 3$ bound entangled state \cite{bennett}, \be\label{3x3}
\rho=\frac{1}{4}(I_{9}-\sum\limits_{i=0}^{4}|\xi_{i}\ra\la\xi_{i}|),
\ee where $I_9$ is the $9\times 9$ identity matrix, $|\xi_{0}\ra
=\frac{1}{\sqrt{2}}|0\ra(|0\ra-|1\ra)$, $|\xi_{1}\ra =
\frac{1}{\sqrt{2}}(|0\ra-|1\ra)|2\ra$, $|\xi_{2}\ra =
\frac{1}{\sqrt{2}}|2\ra(|1\ra-|2\ra)$, $|\xi_{3}\ra =
\frac{1}{\sqrt{2}}(|1\ra-|2\ra)|0\ra$, $|\xi_{4}\ra =
\frac{1}{3}(|0\ra+|1\ra+|2\ra)(|0\ra+|1\ra+|2\ra)$. We simply choose
the local orthonormal observables to be the normalized generators of
$SU(3)$. Formula $(\ref{con0})$ gives $C(\rho)\geq 0.050$. Formula
$(\ref{con})$ gives $C(\rho)\geq 0.052$ \cite{vicente}, while
formula $(\ref{con2})$ yields a better lower bound $C(\rho)\geq
0.0555$.

If we mix the bound entangled state (\ref{3x3}) with
$|\psi\ra=\frac{1}{\sqrt{3}}\sum\limits_{i=0}^{2}|ii\ra$,
$\rho^{'}=(1-x)\rho+x|\psi\ra\la\psi|$,
it is easily seen that $(\ref{con2})$ gives a
better lower bound of concurrence than formula $(\ref{con0})$ (Fig. 1).

\begin{figure}[tbp]
\begin{center}
\resizebox{10cm}{!}{\includegraphics[bb=0 0 430 230]{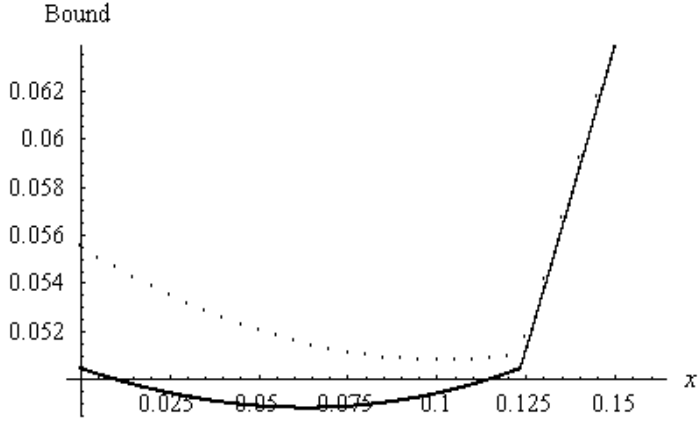}}
\end{center}
\caption{Lower bounds from (12) (dashed line) and (10) (solid line)}
\label{fig2}
\end{figure}

The separability criteria based on covariance matrix approach
can be generalized to multipartite systems.
We first consider the tripartite case, $\rho_{ABC}\in {\mathcal
{H}}^{A}_{d}\otimes{\mathcal {H}}^{B}_{d}\otimes{\mathcal
{H}}^{C}_{d}$. Take
$d^{2}$ observables $A_{k}$ on ${\mathcal {H}}_{A}$ resp. $B_{k}$ on
${\mathcal {H}}_{B}$ resp. $C_{k}$ on ${\mathcal {H}}_{C}$. Set
$\{M_{k}\}=\{A_{k}\otimes I\otimes I, I\otimes B_{k}\otimes I,
I\otimes I \otimes C_{k}\}$.
The covariance matrix defined by $(\ref{gammae})$ has then the
following block structure:
\begin{eqnarray}\label{d3}
\gamma=\left(%
    \begin{array}{ccc}
      A & D & E \\
      D^{T} & B & F \\
      E^{T} & F^{T} & C \\
    \end{array}
    \right),
\end{eqnarray}
where $A=\gamma(\rho_{A},\{A_{k}\})$, $B=\gamma(\rho_{B},\{B_{k}\})$,
$C=\gamma(\rho_{C},\{C_{k}\})$, $D_{ij}=\langle A_{i}\otimes B_{j}
\rangle_{\rho_{AB}}-\langle A_{i} \rangle_{\rho_{A}}\langle B_{j}
\rangle_{\rho_{B}}$,
$E_{ij}=\langle A_{i}\otimes C_{j} \rangle_{\rho_{AC}}-\langle A_{i}
\rangle_{\rho_{A}}\langle C_{j} \rangle_{\rho_{C}}$, $F_{ij}=\langle
B_{i}\otimes C_{j} \rangle_{\rho_{BC}}-\langle B_{i}
\rangle_{\rho_{B}}\langle C_{j} \rangle_{\rho_{C}}$.

{\sf [Theorem 2]} If $\rho_{ABC}$ is fully separable, then
\begin{eqnarray}
\label{t112} ||D||_{HS}^{2}&\leq&
(1-tr(\rho_{A}^{2}))(1-tr(\rho_{B}^{2})),\\
\label{t113} ||E||_{HS}^{2}&\leq&
(1-tr(\rho_{A}^{2}))(1-tr(\rho_{C}^{2})),\\
\label{t123} ||F||_{HS}^{2}&\leq&
(1-tr(\rho_{B}^{2}))(1-tr(\rho_{C}^{2})),
\end{eqnarray}
and
\begin{eqnarray}
\label{t212}2||D||_{KF}\leq (1-tr(\rho_{A}^{2}))+(1-tr(\rho_{B}^{2})),\\
\label{t213}2||E||_{KF}\leq (1-tr(\rho_{A}^{2}))+(1-tr(\rho_{C}^{2})),\\
\label{t223}2||F||_{KF}\leq
(1-tr(\rho_{B}^{2}))+(1-tr(\rho_{C}^{2})).
\end{eqnarray}

{\sf [Proof]} For a tripartite product state
$\rho_{ABC}=\rho_{A}\otimes\rho_{B}\otimes\rho_{C}$, $D$, $E$ and
$F$ in $(\ref{d3})$ are zero. If $\rho_{ABC}$ is fully separable,
then there exist states $|a_{k}\rangle\langle a_{k}|$ in ${\mathcal
{H}}^{A}_{d}$, $|b_{k}\rangle\langle b_{k}|$ in ${\mathcal
{H}}^{B}_{d}$ and $|c_{k}\rangle\langle c_{k}|$ in ${\mathcal
{H}}^{C}_{d}$, and $p_{k}$ such that $\gamma(\rho)\geq
\kappa_{A}\oplus \kappa_{B} \oplus \kappa_{C}$, where
$\kappa_{A}=\sum p_{k}\gamma(|a_{k}\rangle\langle
a_{k}|,\{A_{k}\})$, $\kappa_{B}=\sum
p_{k}\gamma(|b_{k}\rangle\langle b_{k}|,\{B_{k}\})$ and
$\kappa_{C}=\sum p_{k}\gamma(|c_{k}\rangle\langle
c_{k}|,\{C_{k}\})$,
 i.e.
\begin{eqnarray}\label{th3}
Y\equiv\left(%
    \begin{array}{ccc}
      A-\kappa_{A} & D & E \\
      D^{T} & B-\kappa_{B} & F \\
      E^{T} & F^{T} & C-\kappa_{C} \\
    \end{array}%
    \right)\geq 0.
\end{eqnarray}
Thus all the $2 \times 2$ minor submatrices of Y must be positive.
Selecting one with two rows and columns from the first two block rows
and columns of Y, we have
\be\label{lll}
\left| \begin{array}{cc}
(A-\kappa_{A})_{ii} & D_{ij}\\
D_{ji} & (B-\kappa_{B})_{jj}\\
\end{array}
\right|\geq 0,
\ee
i.e. $(A-\kappa_{A})_{ii}(B-\kappa_{B})_{jj}\geq |D_{ij}|^{2}$. Summing
over all i, j and using ($\ref{lemma1}$), we get
\begin{eqnarray*}
||D||_{HS}^{2}&=&\sum\limits_{i,j=1}^{d^{2}}D_{i,j}^{2}\leq(tr A-tr
\kappa_{A}))(tr B-tr \kappa_{B})\\
&=&(d-tr(\rho_{A}^{2})-d+1)(d-tr(\rho_{B}^{2})-d+1)
=(1-tr(\rho_{A}^{2}))(1-tr(\rho_{B}^{2})),
\end{eqnarray*}
which proves (\ref{t112}). (\ref{t113}) and (\ref{t123}) can be similarly proved.

From (\ref{lll}) we also have
$(A-\kappa_{A})_{ii}+(B-\kappa_{B})_{ii}\geq 2|D_{ii}|$.
Therefore
\begin{eqnarray}\label{th3p}
\sum\limits_{i}|D_{ii}|&\leq&\frac{(tr A-tr \kappa_{A}))+(tr B-tr
\kappa_{B})}{2} \nonumber\\
&=&\frac{(d-tr(\rho_{A}^{2})-d+1)+(d-tr(\rho_{B}^{2})-d+1)}{2}\nonumber \\
 &=&\frac{(1-tr(\rho_{A}^{2}))+(1-tr(\rho_{B}^{2}))}{2}.
\end{eqnarray}
Note that
$\sum\limits_{i=1}^{d^{2}}D_{ii}\leq\sum\limits_{i=1}^{d^{2}}|D_{ii}|$.
By using that $Tr(MU) \leq||M||_{KF}=Tr \sqrt{MM^{\dag}}$ for any
matrix $M$ and any unitary $U$ \cite{matrix}, we have
$\sum\limits_{i=1}^{d^{2}}D_{ii}\leq ||D||_{KF}$.

Let $D=U^{\dag}\Lambda V$ be the singular value
decomposition of $D$.
Make a transformation of the orthonormal normalized basis of the local orthonormal
observable space: ${\widetilde{A}}_{i}= \sum\limits_{l}U_{il}A_{l}$ and
${\widetilde{B}}_{j}= \sum\limits_{m}V_{jm}^{*}B_{m}$.
In the new basis we have
\be\label{ppp8}
{\widetilde{D}}_{ij}=\sum\limits_{lm}U_{il}D_{lm}V_{jm}=(UDV^{\dag})_{ij}=\Lambda_{ij}.
\ee
Then (\ref{th3p}) becomes
\begin{eqnarray*}
\sum\limits_{i=1}^{d^{2}}\widetilde{D}_{ii}=||D||_{KF}\leq
\frac{(1-tr(\rho_{A}^{2}))+(1-tr(\rho_{B}^{2}))}{2}
\end{eqnarray*}
which proves (\ref{t212}). (\ref{t213}) and (\ref{t223}) can similarly treated.
\hfill \rule{1ex}{1ex}

We consider now the case that $\rho_{ABC}$ is bi-partite
separable.

{\sf [Theorem 3]} If $\rho_{ABC}$ is a bi-partite
separable state with respect to the bipartite
partition of the sub-systems $A$ and $BC$ (resp. $AB$ and $C$; resp.
$AC$ and $B$), then
$(\ref{t112})$, $(\ref{t113})$ and $(\ref{t212})$, $(\ref{t213})$
(resp. $(\ref{t113})$, $(\ref{t123})$ and $(\ref{t213})$, $(\ref{t223})$; resp.
$(\ref{t112})$, $(\ref{t123})$ and $(\ref{t212})$, $(\ref{t223})$)
must hold.

{\sf [Proof]} We prove the case that $\rho_{ABC}$ is bi-partite
separable with respect to the $A$ system and $BC$ systems partition. The other cases
can be similarly treated. In this case the matrices $D$ and $E$ in the covariance
matrix $(\ref{d3})$ are zero. $\rho_{ABC}$ takes the form
$\rho_{ABC}=\sum\limits_{m}p_{m}\rho_{A}^{m}\otimes\rho_{BC}^{m}$.
Define $\kappa_{A}=\sum
p_{m}\gamma(\rho_{A}^{m},\{A_{k}\})$, $\kappa_{BC}=\sum
p_{m}\gamma(\rho_{BC}^{m},\{B_{k}\otimes I, I\otimes C_{k}\})$. $\kappa_{BC}$
has a form
\begin{eqnarray*}
\kappa_{BC}=\left(%
    \begin{array}{cc}
      \kappa_{B} & F^{'} \\
      (F^{'})^{T} & \kappa_{C} \\
    \end{array}%
    \right),
\end{eqnarray*}
where $\kappa_{B}=\sum p_{k}\gamma(|b_{k}\rangle\langle
b_{k}|,\{B_{k}\})$ and $\kappa_{C}=\sum p_{k}\gamma(|c_{k}\rangle\langle
c_{k}|,\{C_{k}\})$,
$(F^{'})_{ij}=\sum\limits_{m}p_{m}(\langle B_{i}\otimes
C_{j}\rangle_{\rho_{BC}^{m}}-\langle
B_{i}\rangle_{\rho_{B}^{m}}\langle C_{j}\rangle_{\rho_{C}^{m}}).$
By using the concavity of covariance matrix we have
\begin{eqnarray*}
\gamma(\rho_{ABC})\geq\sum\limits_{m}p_{m}\gamma(\rho_{A}^{m}\otimes\rho_{BC}^{m})
=\left(%
    \begin{array}{ccc}
      \kappa_{A} & 0 & 0 \\
      0 & \kappa_{B} & F^{'} \\
      0 & (F^{'})^{T} & \kappa_{C} \\
    \end{array}
    \right).
\end{eqnarray*}
Accounting to the method used in proving Theorem 2, we get
$(\ref{t112})$, $(\ref{t113})$ and $(\ref{t212})$, $(\ref{t213})$.
\hfill \rule{1ex}{1ex}

From Theorem 2 and 3 we have

{\sf [corollary]} If two of the inequalities $(\ref{t112})$,
$(\ref{t113})$ and $(\ref{t123})$ (or $(\ref{t212})$, $(\ref{t213})$
and $(\ref{t223})$) are violated, the state must be fully entangled.

The result of Theorem 2 can be generalized to general multipartite case
$\rho\in {\mathcal {H}}_{d}^{(1)}\otimes{\mathcal {H}}_{d}^{(2)}\otimes \cdots
\otimes{\mathcal {H}}_{d}^{(N)}$.
Define  $\hat{A}^{i}_{\alpha}=I\otimes I\otimes \cdots
\lambda_{\alpha}\otimes I\otimes\cdots\otimes I$, where
$\lambda_{0}=I/d$, $\lambda_{\alpha}$ ($\alpha=1, 2,
\cdots d^{2}-1$) are the normalized generators of $SU(d)$ satisfying
$tr\{\lambda_{\alpha}\lambda_{\beta}\}=\delta_{\alpha\beta}$ and acting
on the $i^{th}$ system ${\mathcal {H}}_{d}^{(i)}$, $i=1, 2, \cdots, N$. Denote $\{M_{k}\}$
the set of all $\hat{A}^i_{\alpha}$.
Then the covariance matrix of $\rho$ can be written as
\begin{eqnarray}
\label{dn}
\gamma(\rho)=\left(%
    \begin{array}{ccc}
      {\mathcal {A}}_{11} & {\mathcal {A}}_{12} \cdots & {\mathcal {A}}_{1N} \\
      {\mathcal {A}}_{12}^{T} & {\mathcal {A}}_{22} \cdots & {\mathcal {A}}_{2N} \\
      \vdots      & \vdots             & \vdots      \\
      {\mathcal {A}}_{1N}^{T} & {\mathcal {A}}_{2N}^{T} \cdots & {\mathcal {A}}_{NN} \\
    \end{array}%
    \right),
\end{eqnarray}
where ${\mathcal {A}}_{ii}=\gamma(\rho, \{\hat{A}^{i}_{k}\})$ and
$({\mathcal
{A}}_{ij})_{mn}=\langle\hat{A}^{i}_{m}\otimes\hat{A}^{j}_{n}\rangle-\langle\hat{A}^{i}_{m}
\rangle\langle\hat{A}^{j}_{n}\rangle$ for $i\neq j$.

For a product state $\rho_{12\cdots N}$,
${\mathcal {A}}_{ij}$, $i\neq j$, in $(\ref{dn})$ are zero matrices.
For a fully separable mixed state, it has the form
$\rho=\sum\limits_{k}p_{k}|\psi^{1}_{k}\rangle\langle\psi^{1}_{k}|
\otimes |\psi^{2}_{k}\rangle\langle\psi^{2}_{k}|\otimes \cdots
\otimes |\psi^{N}_{k}\rangle\langle\psi^{N}_{k}|$.
Define
\begin{eqnarray}
\kappa_{{\mathcal
{A}}_{ii}}=\sum\limits_{k}p_{k}\gamma(|\psi^{i}_{k}\rangle\langle\psi^{i}_{k}|,
\{\hat{A}^{i}_{l}\}).
\end{eqnarray}
Then for a fully separable multipartite state $\rho$ one has
\begin{eqnarray}
Z=\left(%
    \begin{array}{ccc}
      {\mathcal {A}}_{11}-\kappa_{{\mathcal {A}}_{11}} & {\mathcal {A}}_{12} \cdots & {\mathcal {A}}_{1N} \\
      {\mathcal {A}}_{12}^{T} & {\mathcal {A}}_{22}-\kappa_{{\mathcal {A}}_{22}} \cdots & {\mathcal {A}}_{2N} \\
      \vdots      & \vdots             & \vdots      \\
      {\mathcal {A}}_{1N}^{T} & {\mathcal {A}}_{2N}^{T} \cdots & {\mathcal {A}}_{NN}-\kappa_{{\mathcal {A}}_{NN}} \\
    \end{array}%
    \right)\geq 0.
\end{eqnarray}
From which we have the following separability criterion for multipartite systems:

{\sf[Theorem 4]} If a state $\rho\in {\mathcal
{H}}_{d}^{(1)}\otimes{\mathcal {H}}_{d}^{(2)}\otimes \cdots
\otimes{\mathcal {H}}_{d}^{(N)}$ is fully separable, the following
inequalities
\begin{eqnarray}
||{\mathcal {A}}_{ij}||_{HS}^{2}&&\leq (1-tr(\rho_{i}^{2}))(1-tr(\rho_{j}^{2})),\\[3mm]
||{\mathcal {A}}_{ij}||_{KF}&&\leq
\frac{(1-tr(\rho_{i}^{2}))+(1-tr(\rho_{j}^{2}))}{2}
\end{eqnarray}
must be fulfilled for any $i\neq j$.

\bigskip

We have studied the separability of quantum states by using the
covariance matrix. An alternative separability criterion
has obtained for bipartite systems, which is a
supplement of the criterion in \cite{0611282}.
The covariance matrix approach has been applied to multipartite systems
and some related separability criteria have been obtained.
The local orthonormal observable dependent lower bound of concurrence
proposed in \cite{0611282} has been optimized.

In dealing with the multipartite cases,
we have considered that all subsystems have the same
dimensions. The results can be generalized to the case that
some or all subsystems have different dimensions.
For instance, let $N_{max}=N_{n}$ be the the largest dimension.
One can choose $N_{n}^{2}$ observables $\hat{A}_{k}^{n}$.
For other subsystems with smaller dimensions, say $N_{1}$,
one chooses $N_{1}^{2}$ observables $\hat{A}_{k}^{1}$,
$k=1,\cdots,N_{1}^{2}$,
and set $\hat{A}_{k}^{1}=0$ for $k=N_{1}^{2}+1, \cdots, N_{n}^{2}$.

\bigskip
\noindent{\bf Acknowledgments}\, This work is supported by the NSFC
10675086, NSF of Beijing 1042004 and KM200510028022, NKBRPC(2004CB318000).

\end{document}